\documentclass{article}

\PassOptionsToPackage{numbers, compress}{natbib}
\usepackage[preprint]{neurips_2025}

\usepackage[utf8]{inputenc} 
\usepackage[T1]{fontenc}    
\usepackage{hyperref}       
\usepackage{url}            
\usepackage{booktabs}       
\usepackage{amsfonts}       
\usepackage{nicefrac}       
\usepackage{microtype}      
\usepackage{xcolor}         
\usepackage{booktabs} 
\usepackage{svg}

\usepackage{graphicx}   
\usepackage{caption}    
\usepackage{arydshln} 
\usepackage[table]{xcolor}

\title{Asm2SrcEval: Evaluating Large Language Models for Assembly-to-Source Code Translation}

\author{
  Parisa Hamedi,
  Hamed Jelodar,
  Samita Bai,
  Mohammad Meymani,\\
  \textbf{Roozbeh Razavi-Far},
  \textbf{Ali Ghorbani}\\
  Canadian Institute for Cybersecurity\\
  University of New Brunswick, Faculty of Computer Science\\
  \{parisa.hamedi,h.jelodar,samita.bai,mohammad.meymani79,\\roozbeh.razavi-far,ghorbani\}@unb.ca
}

\begin{document}

\maketitle

\begin{abstract}
  Assembly-to-source code translation is a critical task in reverse engineering, cybersecurity, and software maintenance, yet systematic benchmarks for evaluating large language models on this problem remain scarce. In this work, we present the first comprehensive evaluation of five state-of-the-art large language models on assembly-to-source translation. We assess model performance using a diverse set of metrics capturing lexical similarity (BLEU, ROUGE, and METEOR), semantic alignment (BERTScore), fluency (Perplexity), and efficiency (time prediction). Our results reveal clear trade-offs: while certain models excel in text similarity metrics, others demonstrate lower perplexity or faster inference times. We further provide qualitative analyses of typical model successes and failure cases, highlighting challenges such as control flow recovery and identifier reconstruction. Taken together, our benchmark offers actionable insights into the strengths and limitations of current large language models for program translation, establishing a foundation for future research in combining accuracy with efficiency for real-world applications.
\end{abstract}

\section{INTRODUCTION}
Assembly language, while powerful, presents significant challenges due to its machine-like syntax, lack of abstractions, and hardware dependence \cite{muchnick1997advanced}. Programs written in assembly are hard to read, maintain, and debug, making development slow and error-prone \cite{mehler2006challenges, wu2019analysis, alfred2007compilers}. Unlike high-level languages such as C or C++, which provide portability, modularity, and human-readable syntax \cite{stroustrup1999overview}, assembly requires deep hardware knowledge and is only practical for small-scale programs. These limitations in readability, scalability, and collaboration (Table \ref{tab:assembly_vs_highlevel}) strongly motivate research into methods that can translate assembly code into higher-level, and more understandable representations.
\begin{table}[h!]
\centering
\caption{Comparison between assembly (low-level) and high-level (C++) programming languages.}
\begin{tabular}{p{3cm} p{5cm} p{5cm}}
\toprule
\textbf{Aspect} & \textbf{Assembly (Low-Level)} & \textbf{High-Level (C++ etc.)} \\
\midrule
Readability     & Hard to read (machine-like, cryptic instructions) & Human-readable, closer to natural language \\
Maintainability & Very difficult to modify/debug & Easier to maintain and refactor \\
Portability     & Hardware-specific, not portable across CPUs & Portable across architectures via compilers \\
Development Speed & Slow (manual registers, memory management) & Faster (loops, functions, data structures) \\
Error-Proneness & Easy to introduce subtle bugs & Safer abstractions, type-checking reduces errors \\
Abstractions    & No functions, classes, or advanced data types & Supports OOP, modularity, libraries \\
Scalability     & Only feasible for small programs & Suitable for large, complex software systems \\
Collaboration   & Requires deep hardware knowledge (few can contribute) & Accessible to more developers, teamwork-friendly \\
\bottomrule
\end{tabular}
\label{tab:assembly_vs_highlevel}
\end{table}

Previous studies have investigated assembly-to-source translation using rule-based or statistical approaches \cite{yakdan2015no, NEURIPS2019_093b60fd, fu2019coda, cao2022boosting, hosseini2022beyond}. While these efforts demonstrated the feasibility of the task, they often relied on handcrafted rules or shallow machine learning methods \cite{lacomis2019dire, nitin2021direct, chen2022augmenting}, and, thus, struggled to capture semantic nuances across diverse assembly instructions \cite{dramko2024taxonomy, cao2024evaluating, zhou2025decompiling}.

Recent advances in large language models (LLMs) have shown strong capabilities in many domains including source-to-source translation, code synthesis, code-to-natural-language-description, and natural language-to-code generation, making them natural candidates for tackling the challenges of assembly-to-source translation \cite{jelodar2025large,jelodar2025large2,jelodar2025nld}. Building on these strengths, several works have begun to explore LLMs for decompilation. For example, LLM4Decompile \cite{tan2024llm4decompile} demonstrated converting binaries into readable source code, \cite{wong2023refining} improved recompilability of decompiler outputs, and Decompile-Bench \cite{tan2025decompile} introduced large-scale benchmarks to facilitate systematic evaluation. However, despite these advances, no work has specifically examined direct LLM-based translation from assembly to C++ code, leaving this as an open and practically significant research problem.

In this work, we present a benchmark to evaluate five representative LLMs on the task of translating assembly code into high-level source code (C++). We assess their performance using widely adopted automatic evaluation metrics, including BLEU, ROUGE, METEOR, BERTScore, Perplexity, and prediction time \cite{hu2024unveiling, brown2024enhancing}. Our findings reveal notable trade-offs between fluency and correctness, highlighting both the strengths and limitations of current LLMs. These results provide valuable insights into the challenges of assembly-to-source code translation and underscore its practical significance for software engineering, program comprehension, and security analysis.

\section{RELATED WORKS}
Early efforts in assembly-to-source translation primarily relied on rule-based systems and statistical learning approaches. These methods mapped instruction patterns to higher-level constructs through handcrafted heuristics \cite{yakdan2015no, fu2019coda, hosseini2022beyond}. While effective on restricted subsets of assembly, they often lacked scalability and failed to capture the semantic nuances of diverse instruction sets \cite{lacomis2019dire, nitin2021direct, chen2022augmenting}. More recent statistical and neural approaches have sought to improve upon these limitations by introducing learned representations of code, yet they still depend heavily on aligned training data and are brittle to out-of-distribution inputs \cite{dramko2024taxonomy, cao2024evaluating, zhou2025decompiling}.

In parallel, compiler-inspired decompilers and binary analysis frameworks have been developed to reconstruct high-level semantics from low-level code \cite{NEURIPS2019_093b60fd, cao2022boosting}. These tools use static and dynamic analysis techniques to recover control flow, data structures, and variable names, but the resulting code is often verbose, hard to read, or not recompilable, limiting their usefulness for software maintenance and reverse engineering.

The emergence of LLMs has opened a new avenue for program translation. LLMs trained on large-scale code corpora have demonstrated strong performance in code synthesis, translation, and bug fixing \cite{jelodar2025large,jelodar2025large2,hu2024unveiling, brown2024enhancing,meymani2025can}. Several recent studies have begun to apply LLMs to decompilation tasks: LLM4Decompile \cite{tan2024llm4decompile} showed that LLMs can translate binaries into human-readable source code, Wong et al. \cite{wong2023refining} focused on improving the recompilability of generated outputs, and Decompile-Bench \cite{tan2025decompile} introduced a benchmark to standardize evaluation. However, these works do not specifically target direct assembly-to-C++ translation, nor do they provide a systematic comparison across multiple models and evaluation dimensions.

Our work fills this gap by introducing the first benchmark that evaluates a diverse set of LLMs on assembly-to-C++ translation. We assess performance along lexical, semantic, fluency, and efficiency dimensions, offering new insights into the trade-offs between accuracy and practicality in this important task.

\section{METHODOLOGY}
\begin{figure}[h]
\centering
\includegraphics[trim=2cm 2.5cm 6.5cm 2.5cm,clip,width=1\textwidth]{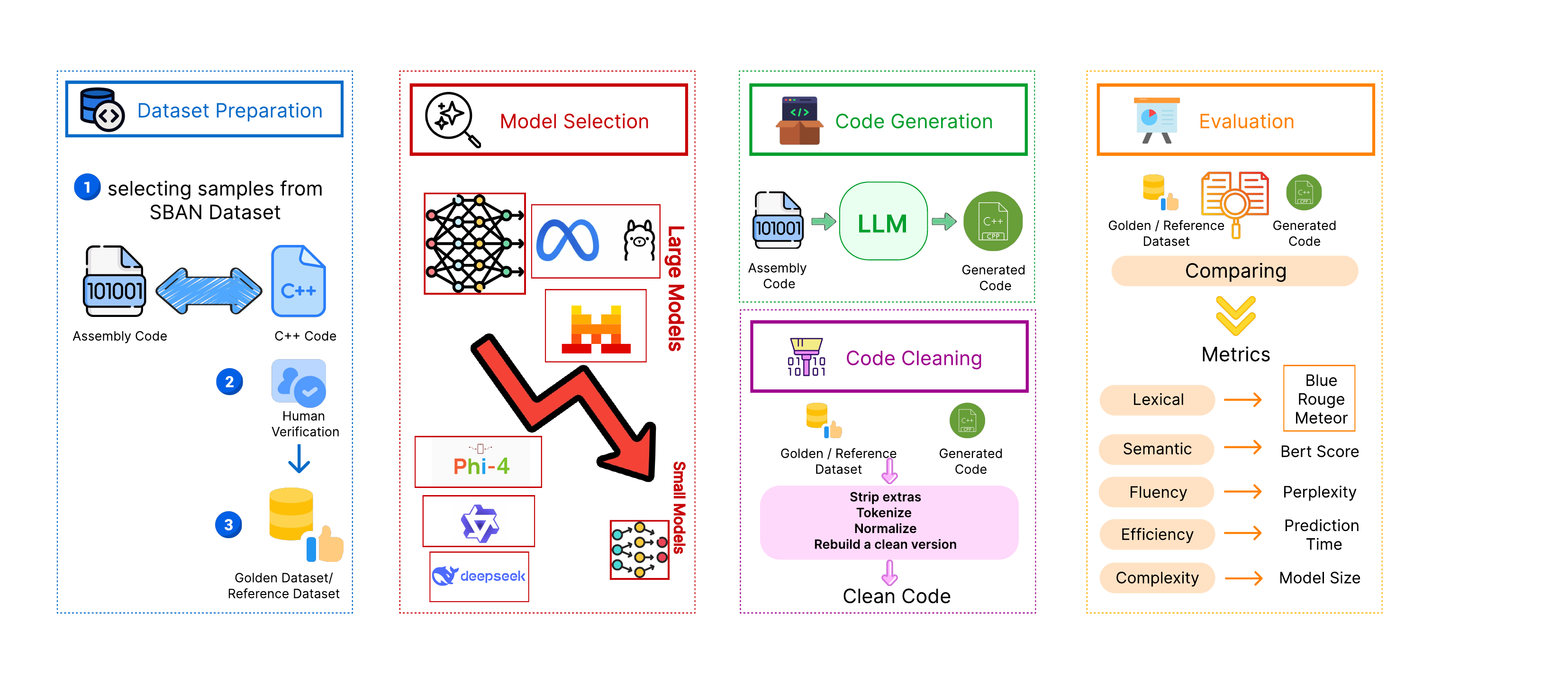}
\caption{The methodology workflow of our proposed approach.}
\label{fig:methodology}
\end{figure}

In this section, we outline the methodology used to evaluate large language models (LLMs) for the task of translating assembly code into C++. Figure \ref{fig:methodology} shows the workflow of our research work, which consists of four phases of data collection, model selection, code generation, and evaluation. First, we provide an overview of the five selected LLMs and highlight their key features, training paradigms, and relevance to code translation. Next, we describe the evaluation metrics—BLEU, ROUGE, METEOR, BERTScore, Perplexity, and prediction time—and explain the specific goal of using each in assessing model performance. We then present the dataset and reference benchmark used in our experiments, including their composition and distinguishing characteristics. Finally, we report and analyze the experimental results obtained across the different models and metrics, offering both quantitative comparisons and qualitative insights.
\subsection{LLM Models}
To evaluate the task of assembly-to-C++ translation, we selected five representative instruction-tuned large language models (LLMs), chosen to cover a range of scales and architectures. The models can be divided into two groups based on their parameter size. The small-scale models include DeepSeek-Coder-1.3B-Instruct \cite{guo2024deepseek}, Phi-4-Mini-Instruct (Microsoft) \cite{abouelenin2025phi}, and Qwen2.5-Coder-1.5B-Instruct \cite{hui2024qwen2}, which are lightweight models (1-2B parameters) designed for efficiency and suitable for deployment in resource-constrained environments. These models are particularly attractive for edge computing and IoT scenarios, where inference speed and memory footprint are critical.

The larger-scale models consist of Llama-3.1-8B-Instruct (Meta) \cite{Grattafiori2024Llama3} and Mistral-7B-Instruct-v0.1 \cite{Jiang2023Mistral7B}, which have significantly more parameters (7-8B) and generally provide stronger reasoning and language generation capabilities. These models, while more computationally demanding, offer higher capacity for capturing long-range dependencies and complex patterns in code.

Across both groups, all models share a common focus on instruction tuning, meaning they are optimized to follow user prompts and generate context-aware responses. However, they differ in their design philosophies: DeepSeek-Coder and Qwen2.5-Coder emphasize code-centric pretraining, Phi-4-Mini is optimized for compact general-purpose reasoning, while Llama-3.1 and Mistral focus on broad multilingual and multi-domain adaptability. This diversity allows us to assess how model size and training specialization affect performance in the assembly-to-source translation task. Table~\ref{tab:LLM_Models} summarizes the key characteristics of the five models, highlighting their size, specialization, and distinctive features.
\begin{table}[h!]
\centering
\caption{Comparison of the selected LLMs used for assembly-to-C++ translation.}
\begin{tabular}{l l l}
\hline
\textbf{Model} & \textbf{Size} & \textbf{Key Features} \\
\hline
DeepSeek-Coder-1.3B-Instruct & 1.3B & Code-focused, instruction-tuned \\
Phi-4-Mini-Instruct & 3.8B & Compact, efficient, general-purpose \\
Qwen2.5-Coder-1.5B-Instruct & 1.5B & Code-centric, multilingual \\
Mistral-7B-Instruct-v0.1 & 7B & General-purpose, strong reasoning \\
Llama-3.1-8B-Instruct & 8B & Broad coverage, multilingual \\
\hline
\end{tabular}
\label{tab:LLM_Models}
\end{table}

\subsection{Evaluation Metrics}
To assess the performance of LLMs on assembly-to-C++ translation, we employ six evaluation metrics: BLEU, ROUGE, METEOR, BERTScore, Perplexity, and Prediction Time. Each captures a distinct dimension of translation quality or practicality, and together they provide a comprehensive evaluation framework.
\begin{itemize}
\item BLEU (Bilingual Evaluation Understudy): Measures n-gram overlap between generated output and the reference code. It primarily evaluates syntactic correctness but may penalize valid variations in phrasing \cite{post2018call, 8813269, reiter2018structured}.

\item ROUGE (Recall-Oriented Understudy for Gisting Evaluation) is a family of n-gram overlap metrics that emphasize recall by rewarding outputs which capture more of the reference content. Although originally developed for natural language evaluation, ROUGE has been widely applied across domains, including code translation. Its main variants include ROUGE-1 (unigram overlap), ROUGE-2 (bigram overlap), ROUGE-L (longest common subsequence), and ROUGE-Lsum (a summarization-oriented extension). Each variant reflects a distinct inductive bias: ROUGE-1 measures token coverage, often yielding high scores even when tokens are scrambled; it is useful for assessing vocabulary correctness but ignores structural fidelity. ROUGE-2 captures local order by evaluating bigram overlap, rewarding adjacent-token correctness but penalizing small edits disproportionately—for example, inserting a modifier can disrupt many bigrams. ROUGE-L, based on longest common subsequence, evaluates global sequence alignment: it rewards preservation of overall token order while tolerating minor insertions or deletions, making it well-suited to distinguish meaningful control-flow consistency from benign stylistic changes. ROUGE-Lsum is a sentence-aware variant designed for summarization, but for code (where sentence boundaries are irrelevant) it closely tracks ROUGE-L without providing additional insights.

        \begin{figure}[ht]
        \centering
        \includegraphics[trim=3.3cm 0cm 3.4cm 0cm,clip,width=0.9\textwidth]{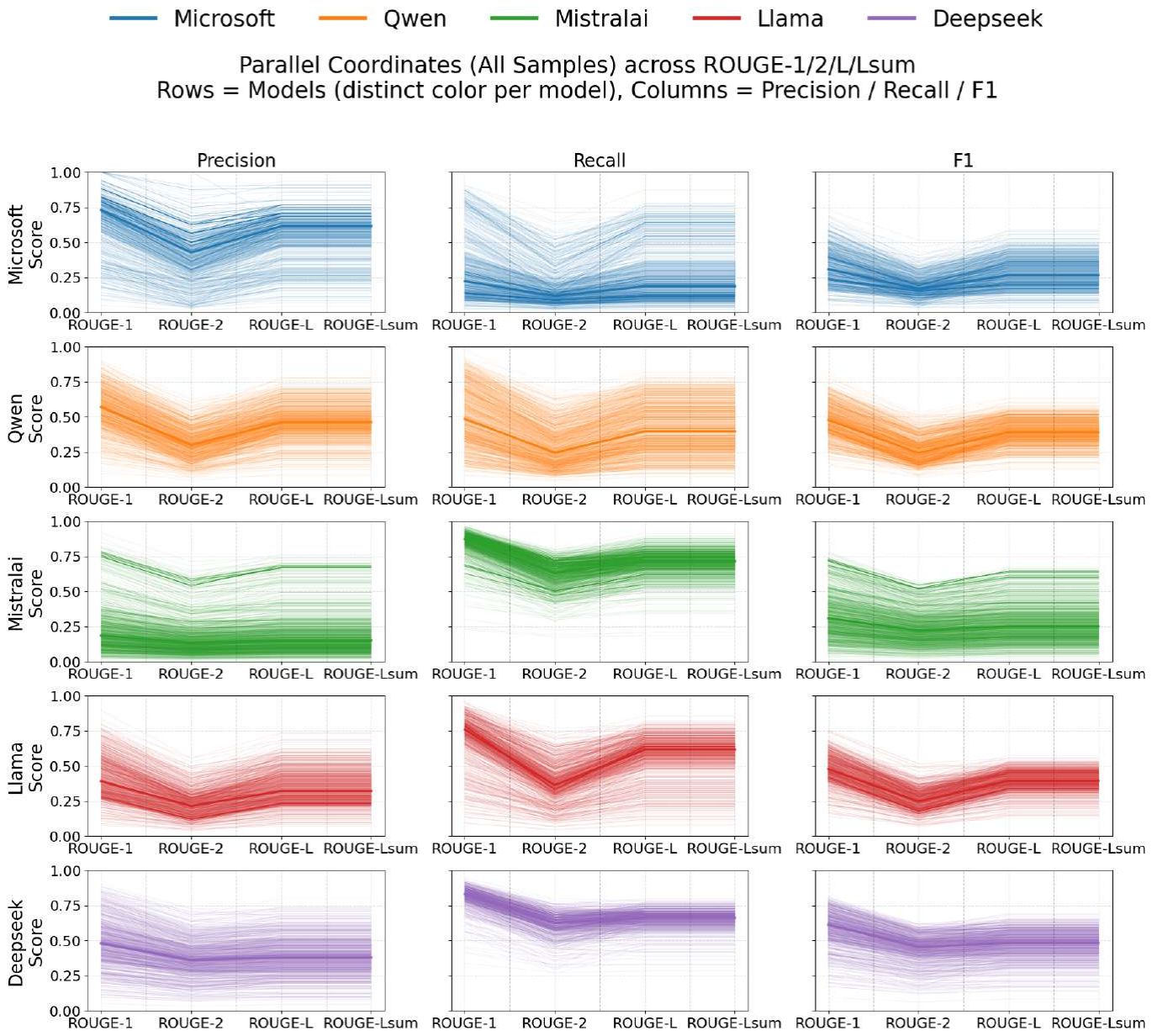}
        \caption{Parallel coordinate plots of five LLMs (Microsoft, Qwen, Mistral, Llama, DeepSeek) across ROUGE metrics (ROUGE-1, ROUGE-2, ROUGE-L, ROUGE-Lsum) for precision, recall, and F1. All models exhibit sharp performance drops on ROUGE-2 compared to ROUGE-1 and ROUGE-L. ROUGE-L and ROUGE-Lsum trends are nearly identical, indicating similar sequence-level and summarization-level matching. Precision generally exceeds recall, particularly for Microsoft and Qwen, suggesting models often produce plausible but incomplete outputs.}
        \label{fig:rougeplot}
      \end{figure}
      
In our parallel-coordinates analysis (Fig. \ref{fig:rougeplot}), all five models exhibit a pronounced trough at ROUGE-2 across Precision, Recall, and F1, reflecting the metric’s brittleness to small local edits common in decompilation (e.g., toggling argument order, inserting casts, or relocating declarations). By contrast, ROUGE-1 often appears inflated: precision is high even when recall or token order is imperfect, since most tokens are present. ROUGE-L produces consistently higher and more stable curves than ROUGE-2 across all models, while ROUGE-Lsum overlaps ROUGE-L, confirming that sentence segmentation offers no additional value in this task. Collectively, these results suggest that sequence-aware yet order-tolerant matching provides the most reliable signal for assembly-to-source translation. Accordingly, we adopt ROUGE-L F1 as our primary ROUGE metric, since it balances precision (avoiding hallucinated tokens) and recall (capturing required tokens), while its LCS foundation preserves structural alignment—an aspect most closely aligned with compilable, semantically faithful code translations. For completeness, we also report ROUGE-1 and ROUGE-2 in the appendix.

\item METEOR (Metric for Evaluation of Translation with Explicit ORdering): Incorporates stemming, synonyms, and word order, making it more sensitive to semantic equivalence compared to BLEU or ROUGE\cite{denkowski2014meteor, banerjee2005meteor}.

\item BERTScore: Uses contextual embeddings from pretrained transformers to measure semantic similarity. It can capture meaning preservation even when surface tokens differ\cite{hanna2021fine, zhang2019bertscore, sun2022bertscore}.
\begin{figure}[ht]
        \centering
        \includegraphics[trim=0cm 8.5cm 0cm 8.5cm,clip,width=0.9\textwidth]{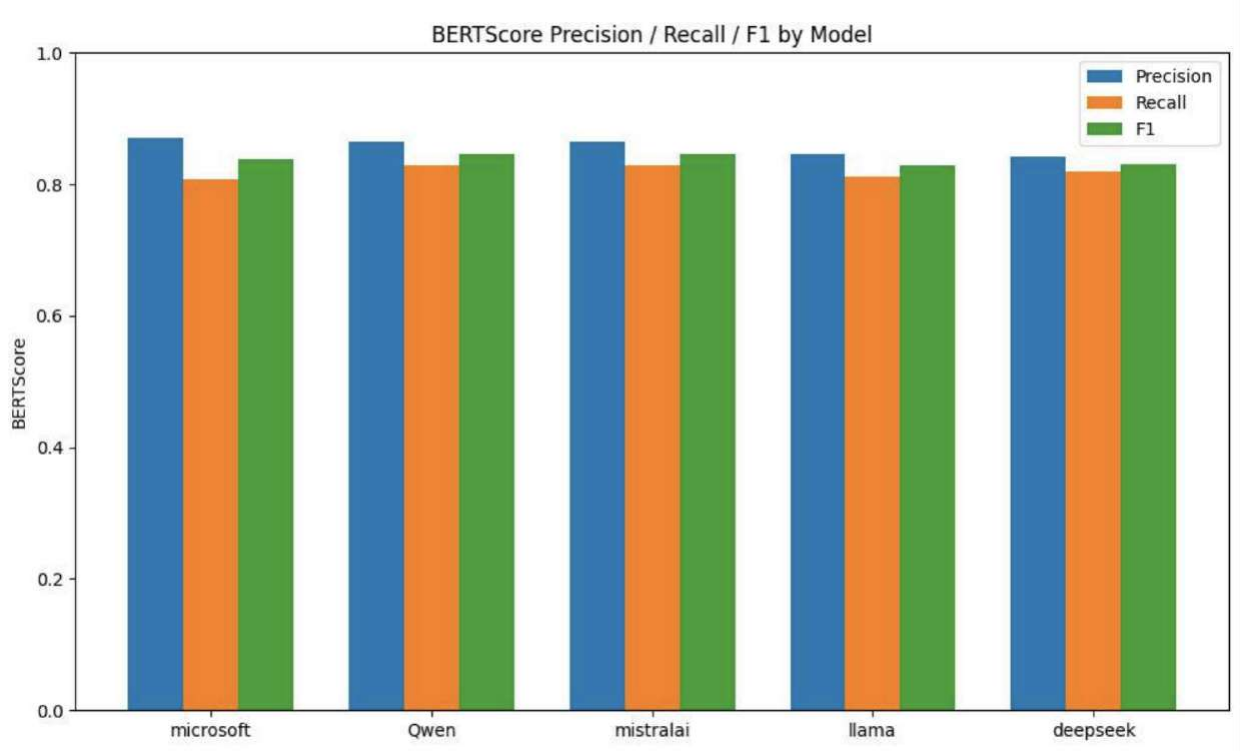}
        \caption{BERTScore evaluation (Precision, Recall, and F1) of five LLMs (Microsoft, Qwen, MistralAI, LLaMA, and DeepSeek) on the assembly-to-source code translation task.}
        \label{fig:bertscore}
      \end{figure}
Figure \ref{fig:bertscore} presents the BERTScore evaluation results (Precision, Recall, and F1) for five large language models—Microsoft, Qwen, MistralAI, LLaMA, and DeepSeek—on the Assembly-to-Source Code Translation task. BERTScore measures semantic similarity between the generated and reference code using contextual embeddings rather than relying solely on surface-level token overlap. As shown in the bar chart, all models achieve relatively high scores across the three metrics, with Precision typically slightly higher than Recall. For reporting purposes, the F1 score is selected as the representative metric since it balances both Precision (exactness) and Recall (coverage). This makes F1 a more robust indicator of overall performance, particularly in translation tasks where both accurate token generation and comprehensive coverage of the reference meaning are equally important.

\item Perplexity: Quantifies fluency and naturalness by measuring how likely the generated sequence is under a model’s probability distribution. Lower perplexity reflects greater confidence and smoother generation \cite{meister2021language, ankner2024perplexed}.

\item Prediction Time: Measures the computational efficiency of producing translations, expressed as the time per output. This metric is particularly important for deployment in real-time or resource-constrained environments, such as edge devices or IoT systems.
\end{itemize}
No single metric fully reflects translation quality and usability. Overlap-based metrics (BLEU, ROUGE, METEOR) capture surface-level correctness, embedding-based BERTScore measures semantic fidelity, perplexity reflects fluency, and prediction time assesses practical feasibility. Together, these metrics balance syntactic correctness, semantic accuracy, fluency, and computational efficiency—dimensions that are all essential for reliable and deployable assembly-to-C++ translation.
\subsection{SBAN Dataset}
For our experiments, we employed a subset of the SBAN dataset \cite{jelodar2025sban}, a benchmark designed for analyzing assembly code. From this collection, we selected about 700 samples, which represents approximately 0.1\% of the full dataset. This subset contains a mix of malware and benign code, reflecting the diversity of real-world assembly programs. Before use, the samples were preprocessed to ensure consistency: formatting was normalized, extraneous metadata was removed, and assembly instructions were paired with their corresponding source-level representations.

To provide a ground truth for evaluation, we used a reference dataset consisting of the C++ source code aligned with the same SBAN assembly samples. This C++ reference code was manually provided and verified by human experts, ensuring semantic correctness and eliminating potential ambiguities. By aligning assembly instructions with trusted source-level counterparts, this dataset enables a reliable assessment of translation quality.


\begin{figure}[h]
        \centering
        \includegraphics[trim=0cm 0cm 0cm 0cm,clip,width=0.9\textwidth]{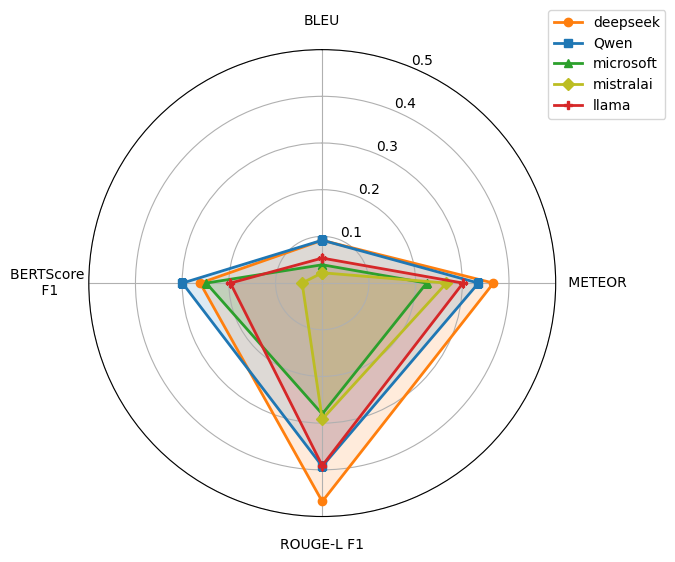}
        \caption{Spider plot comparing models across metrics that are aligned with performance (higher values indicate better results): BLEU, METEOR, ROUGE-L F1, and BERTScore F1. Each axis is min–max normalized across models to enable direct comparison. The plot highlights relative strengths and weaknesses of the models, with overlapping traces indicating identical performance on these metrics.}
        \label{fig:spider}
      \end{figure}

\subsection{Training Settings and Hyperparameters}

All experiments were conducted on an NVIDIA H100 GPU, which provides high memory bandwidth and powerful parallel processing for efficient model training. We trained the models for 10 epochs with a batch size of 32 sequences per GPU, using a maximum sequence length of 512 tokens. The learning rate was set to 3e-5 with a linear warm-up over the first 1,000 steps, and optimization was performed using AdamW ($\beta_1=0.9$, $\beta_2=0.999$, weight decay 0.01) with gradient clipping at 1.0. These hyperparameters were carefully selected to ensure stable training, fast convergence, and optimal performance on our dataset.

\subsection{Results}
We evaluate the five LLMs on six metrics spanning correctness, semantics, fluency, and efficiency. Table 3 summarizes the average results across all models, reporting values for lexical metrics (BLEU, METEOR, ROUGE-L F1), semantic similarity (BERTScore F1), fluency (Perplexity), and efficiency (Prediction Time). Model size is also listed, expressed in billions of parameters, while prediction time is measured in seconds. This table provides a unified view of performance, complementing the visual analyses presented in the spider and bubble charts.

Aligned metrics. The spider plot in Figure \ref{fig:spider} illustrates the aligned metrics—BLEU, METEOR, ROUGE-L F1, and BERTScore F1—where higher values correspond to better performance. Across the evaluated models, Deepseek achieves the highest scores on ROUGE-L F1 and METEOR, indicating strong capability in producing structurally coherent and fluently aligned output. Qwen leads on BERTScore F1, suggesting comparatively better semantic similarity to reference responses. In contrast, mistralai, microsoft, and llama show overall lower performance across most metrics. The close proximity of the curves for Deepseek and Qwen highlights their competitive advantage among the tested models, while the remaining models form a noticeably smaller performance envelope. Overall, the plot emphasizes meaningful separation between higher-performing systems and mid-tier alternatives, with no clear convergence patterns between model sizes.

\begin{figure}[ht]
        \centering
        \includegraphics[trim=0.2cm 0cm 0.2cm 0cm,clip,width=0.9\textwidth]{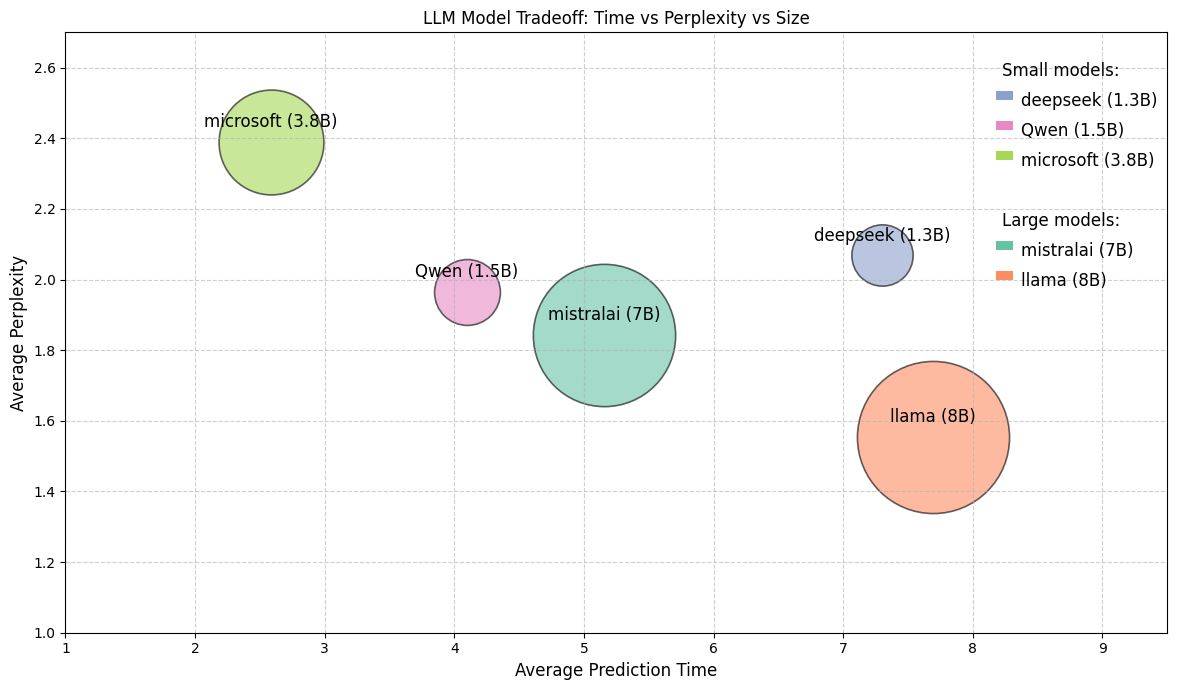}
        \caption{Bubble chart showing the efficiency trade-off across models. X-axis: average prediction time (s). Y-axis: average perplexity (lower is better for both). Bubble area $\propto$  parameter count (billions), color encodes the model. Models closer to the lower-left are more efficient; the legend groups small ($\leq$4B) vs large ($>$4B) models.}
        \label{fig:per_time_size}
      \end{figure}
Non-aligned metrics. In contrast, the bubble chart in Figure \ref{fig:per_time_size} focuses on non-aligned metrics, where lower values are better. Here, the Y-axis denotes perplexity (fluency), the X-axis shows prediction time (efficiency in seconds), and bubble area encodes model size (in billions of parameters). Smaller models, such as Microsoft-Phi-3B, achieve substantially lower prediction times (as fast as 2.59 s), while DeepSeek-1.3B and Qwen-1.5B balance efficiency with competitive perplexity. Large models, particularly Llama-8B, deliver the lowest perplexity (4.44) but incur the highest inference latency (7.69 s), illustrating the cost of scaling.
\begin{table*}[ht]
\centering
\setlength{\tabcolsep}{6pt}
\renewcommand{\arraystretch}{1.2}
\caption{Comparison of models across evaluation metrics. Metrics with ↑ mean higher is better, and ↓ mean lower is better. Highest values are in green, lowest in red. Size is the number of parameters of model in billions (B). Prediction time is in second (s).}
\label{tab:all-metrics}
\begin{tabular}{lccccc}
\toprule
& Microsoft & Qwen & Mistralai & Llama & Deepseek \\
\midrule
BLEU (↑)& 0.0392 & \cellcolor{green!20}0.0918 & \cellcolor{red!20}0.0221 & 0.0535& 0.0916 \\
ROUGE-L F1 (↑)& \cellcolor{red!20}0.2805 & 0.3924 & 0.2915 & 0.3910 & \cellcolor{green!20}0.4678\\
METEOR (↑)& \cellcolor{red!20}0.2236 & 0.3346 & 0.2660 & 0.3015& \cellcolor{green!20}0.3667\\
BERTScore F1 (↑)& 0.2485 & \cellcolor{green!20}0.2994 & \cellcolor{red!20}0.0431 & 0.1967& 0.2615 \\ \\
Perplexity (↓)& 12.9007 & 9.5993 & \cellcolor{red!20}17.9429 & \cellcolor{green!20}4.4414& 10.0233\\
Prediction Time (s) (↓)& \cellcolor{green!20}2.59 s & 4.10 s & 5.16 s & \cellcolor{red!20}7.69 s& 7.3 s\\
Size (B) (↓)& 3.8 B & 1.5 B & 7 B & \cellcolor{red!20}8 B & \cellcolor{green!20}1.3 B \\
\bottomrule
\end{tabular}
\label{tab:allmetrics}
\end{table*}

Overall comparison. Table \ref{tab:allmetrics}, together with Figures \ref{fig:spider} and \ref{fig:per_time_size}, highlights the trade-offs between aligned and non-aligned metrics. Larger models dominate in lexical and semantic similarity but sacrifice inference speed, while smaller models are more efficient yet less consistent on accuracy-based metrics. These findings reveal that model choice depends on the intended application: accuracy-critical tasks benefit from large models, whereas efficiency-critical deployments favor smaller ones. This balance between billions of parameters and seconds of inference time underscores the dual challenge of achieving both correctness and practicality in assembly-to-source translation.

\section{DISCUSSION}
Our evaluation shows that model performance depends strongly on the trade-off between \textbf{accuracy} and \textbf{efficiency}.  

\textbf{Large-scale models.} \emph{Mistral-7B} and \emph{Llama-8B} consistently achieve the best results on aligned metrics (BLEU, ROUGE-L, METEOR, BERTScore), reflecting their superior capacity for semantic and structural fidelity. \emph{Llama-8B} also attains the lowest perplexity (4.44), but its prediction time is the slowest (7.69 s). \emph{Mistral-7B} offers a better balance, combining strong accuracy with more moderate runtime.  

\textbf{Small-scale models.} \emph{Microsoft-Phi-3B} is the fastest (2.59 s) but lags on similarity metrics. \emph{DeepSeek-1.3B} is similarly efficient but weaker overall. \emph{Qwen-1.5B} stands out among small models, achieving competitive BLEU and BERTScore while maintaining reasonable prediction time (4.10 s).  

\textbf{Best models by use case.} For accuracy-critical tasks (e.g., reverse engineering), \emph{Mistral-7B} is the most practical, with \emph{Llama-8B} providing the strongest fluency at higher cost. For efficiency-focused applications (e.g., real-time security analysis), \emph{Microsoft-Phi-3.8B} is preferable. \emph{Qwen-1.5B} represents the best compromise between the two extremes.  

\textbf{Takeaway.} There is no single best model: large models maximize fidelity, while small models enable faster and more resource-conscious deployment. Future work should explore hybrid strategies (e.g., distillation, ensembles) to reduce this trade-off.  

Despite providing new insights, this study faces several limitations. First, our evaluation is conducted on a relatively small subset of the SBAN dataset (~700 samples), which represents only a fraction of the diversity found in real-world assembly code. This restricted scale may limit the generalizability of our findings to broader domains such as obfuscated binaries or domain-specific instruction sets. Second, while we evaluate multiple dimensions of performance—including lexical similarity, semantics, fluency, and efficiency—our benchmark does not incorporate functional correctness checks such as recompilability or execution-based validation. Finally, we only benchmarked five instruction-tuned LLMs; extending the evaluation to larger and more diverse model families would provide a more comprehensive understanding of the trade-offs between accuracy and efficiency in assembly-to-source translation.

\section{CONCLUSION}
This work introduces the first comprehensive benchmark of large language models for assembly-to-C++ translation, evaluating five diverse models across lexical, semantic, fluency, and efficiency metrics. The study reveals trade-offs between semantic accuracy and inference speed, with larger models like Mistral-7B and Llama-8B excelling in fidelity, while smaller ones like Phi-3B and Qwen-1.5B offer faster, more deployment-friendly performance. Based on several samples from the SBAN dataset, the results mark a meaningful starting point but underscore the need for broader evaluations involving more diverse data and additional models. Future directions include exploring hybrid approaches (e.g., distillation, ensembles), incorporating functional correctness via human and compiler feedback, and expanding benchmarks to cover multilingual or domain-specific binaries. Overall, the study lays essential groundwork for advancing LLM-based program translation toward more robust, efficient, and real-world-applicable solutions.

\bibliographystyle{unsrt}  
\bibliography{main}  
\end{document}